\documentclass[journal]{IEEEtran}
\usepackage{amsmath}
\usepackage{amsthm}
\usepackage{amssymb}
\usepackage{cite}
\usepackage{color}
\usepackage[dvips]{graphicx}

\begin{document}
\title{Possible Indications of Electronic Inhomogeneities in Superconducting Nanowire Detectors}
\author{\IEEEauthorblockN{H.L. Hortensius, E.F.C. Driessen, and T.M. Klapwijk\\}
\IEEEauthorblockA{Kavli Institute of Nanoscience, Delft University of Technology, Delft,
The Netherlands\\
Email: h.l.hortensius@tudelft.nl}}

\maketitle

\begin{abstract}
The voltage-carrying state of superconducting NbTiN nanowires, used for single-photon detectors, is analyzed.  Upon lowering the current, the wire returns to the superconducting state in a steplike pattern, which differs from sample to sample. Elimination of geometrical inhomogeneities, such as sharp corners, does not remove these steplike features. They appear to be intrinsic to the material. Since the material is strongly disordered, electronic inhomogeneities are considered as a possible cause.  A thermal model, taking into account random variations of the electronic properties along the wire, is used as an interpretative framework.
\end{abstract}

\section{Introduction}

Superconducting nanowire single-photon detectors (SNSPD's) are promising devices for the detection of single photons because of their fast response time, broadband sensitivity, and low dark-count rate\cite{Natarajan}. A variety of materials are currently being used, with the common denominator that their resistivity in the normal state is unusually large, consistent with an electronic mean free path in the order of the interatomic distance. These materials are chosen for their high critical temperature and fast electron-phonon scattering. In view of the increased interest, the need for reproducible device-fabrication has become an issue. It has been found that often variations of detection efficiency occur from device to device\cite{Kerman} and as a function of position on a single device\cite{Rosticher}, which are not easily understood. Part of this lack of control may signal the need for improved materials control on an atomic level. However, since SNSPD's are  made from highly resistive superconductors, an intrinsic cause might be present as well.

\begin{figure}
\includegraphics{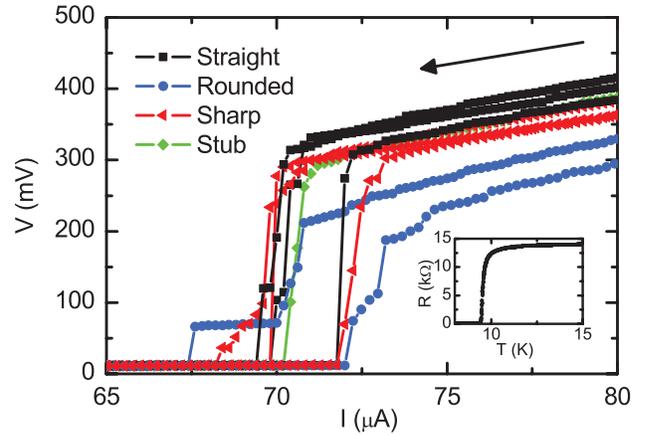}
\caption{Examples of the steplike decrease in voltage observed for decreasing current-bias for various wire geometries. The inset shows the temperature dependence of the resistance of one of the 20~$\mu$m-long, 1~$\mu$m-wide, and 8~nm-thick wires, including the gradually widening contact pads\cite{CurrCrowd}.}
\label{VI_currcrow}
\end{figure}

Due to the short elastic-scattering length, localization is competing with superconductivity, and a tendency to an insulating state is accompanied by a tendency to become superconducting. It is predicted that intrinsic electronic inhomogeneities are formed irrespective of the specific atomic inhomogeneity\cite{Ghosal}. Recent experimental work on TiN has clearly demonstrated the occurrence of such electronic inhomogeneities by measuring the local values of the superconducting energy-gap by scanning tunneling microscopy\cite{Sacepe}. Stimulated by these observations, we have recently studied the microwave electrodynamics of such films\cite{Driessen}, and analyzed the data in the context of a recent theory proposed by Feigel'man and Skvortsov\cite{Feigelman}. Unfortunately, for SNSPD's a limited number of parameters is available to characterize them: the critical temperature, the critical current, the resistivity and the resistive transition. It is, however, reasonable to assume that variations of the superconducting gap, as observed by Sac\'ep\'e \emph{et al.} play a role in the observed variation of detection efficiency from device to device and from position to position on the device. In this manuscript, we focus on one of the few available extra sources of information; the return of the device to the superconducting state, the so-called retrapping current (Fig.~\ref{VI_currcrow}), which is known to vary from sample to sample and shows a clear steplike structure.

\section{Experiments}

The measurements described in this paper, were performed on the same NbTiN wires described in \cite{CurrCrowd}, where the reduction of the critical current due to current crowding was studied. The NbTiN films have a thickness $d$ of $\sim$8~nm, a critical temperature $T_\mathrm{c}$ of 9.5~K, and a resistivity $\rho$ of 170~$\mu\mathrm{\Omega cm}$. The wires have a width $w$ of 1~$\mu$m. We have chosen this width, though it is wider than typical SNSPD devices, since wider wires are less sensitive to fabrication defects and show stronger current-crowding effects. Similar results were obtained for samples as narrow as 50~nm. The length $L$ is 20~$\mu$m. Four different wire geometries were studied: straight wires, corners with optimally designed inner curves, corners with sharp inner curves, and a straight line that included a stub extending 1 $\mu$m to one side.

For each of the devices the current-voltage characteristics were measured in a 4.2~K liquid helium dipstick. The samples were mounted on a ceramic PCB that was in direct contact with a copper sample stage with thermally anchored wires, to ensure a temperature equal to the sample-stage. The temperature of the sample stage itself was controlled by a heater element and measured by a temperature sensor embedded in the sample stage. The measurements reported here are all current-biased in a two-point measurement configuration. The wires were driven into the normal state by applying a current above the critical current of the wire. Then we monitored for each sample for decreasing current the return from the normal state into the superconducting state.

\section{Steplike transitions}

Fig.~\ref{VI_currcrow} shows the retrapping characteristics for four geometries. This set of measurements was taken at 4.2~K. In all samples a stepwise transition from the normal state to the superconducting state is observed. The intermediate states between the normal and superconducting state of the wire are stable for long periods of time (measured up to $\sim$10 minutes) if the bias current is kept stable at this point. The specific pattern of a wire is reproducible over multiple measurements. We observe similar stepwise retrapping characteristics in 100~nm wide NbTiN and TiN nanowires. We find that the presence and stability of these steps does not depend on the geometrical shape of the wire. All geometries show stepwise retrapping patterns. This shows that, for these geometries, the presence of a geometrical `constriction' does not dominate the retrapping characteristics. This is clearly different from the superconducting to normal-state transition, where the critical current is significantly suppressed in geometries with sharp features\cite{CurrCrowd}. The available evidence suggests that these disordered superconducting wires have a tendency to become superconducting by forming one or more superconducting domains, which expand in a steplike discontinuous way.

We have recently performed a detailed study of a model system consisting of a superconducting wire between two normal reservoirs. The analysis was executed taking the full non-equilibrium processes into account\cite{Vercruyssen}. In the strongly disordered wires of NbTiN (and TiN) the origin of the steps needs to be determined prior to a more detailed analysis. It has been theoretically suggested that for a superconductor with homogeneous disorder the superconducting state tends to develop mesoscopic fluctuations in the value of the energy gap and a flattening of the peak in the density of states\cite{Ghosal}. Scanning tunneling microscopy measurements on thin films of TiN, with a resistivity even higher than our films: $\rho > 1000\mu\Omega$cm, show such a variation of the gap energy over the film\cite{Sacepe}. The maximum variation of the gap energy is about 20\% of the total gap energy, and the typical size of the regions is about 50~nm, much larger than the grain size of the films. Variations in the geometry of the nanowire, such as thickness variations due to steps in the underlying substrate, might lead to similar behavior as modeled here\cite{Noat}. We use SEM-inspection to guaranty the homogeneity of the width. We observe similar behavior in ALD-deposited TiN nanowires on an amorphous silicon oxide substrate. The persistent presence of these steplike features for wires that appear `perfect', leads us to conjecture that electronic inhomogeneities play a role.

\section{Thermal model with randomly fluctuating parameters} \label{sec:model}

We assume that the transition occurs by the formation of expanding superconducting domains, with the observed resistance given by the backscattering resistance in the remaining normal part of the wire. In principle, for a driven system consisting of superconducting and normal domains one would have to include current conversion processes at the interfaces. Since we are interested in the possible influence of random electronic inhomogeneities, we use the much more tractable thermal model of the superconducting wire introduced by Skocpol, Beasley, and Tinkham \cite{SBT}.

\begin{figure}
\includegraphics[width=8.5cm]{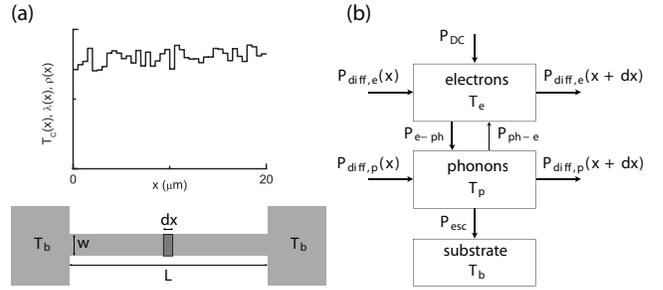}
\caption{Schematic representation of the model. (a) To model inhomogeneities, one can let the critical temperature, heat transfer coefficients, or resistivity vary as a function of position in the wire. (b) In each slice of the wire, the electron and phonon system are assumed to have a well-defined temperature. Energy enters the electron system due to Ohmic dissipation in the normal regions. Heat diffusion along the wire occurs both in the electron and in the phonon system. Heat is exchanged between the electrons and the phonons, and between the phonons and the substrate.}
\label{model}
\end{figure}

In Fig.~\ref{model} we illustrate our model. The bulk contacts are assumed to be in the superconducting state and both the electron- and phonon-temperature are taken to be at the bath temperature of 4.2~K at the contacts (at $x=0$ and $x=L$ in Fig.~\ref{model}(a)). The heat flow model in each segment of length $dx$ is shown in Fig.~\ref{model}(b). We assume that in each slice of the wire the electrons and phonons each have a thermal distribution, with temperatures $T_\mathrm{e}$ and $T_\mathrm{p}$ respectively. The wire is biased with a constant current $I$, and has a normal state resistance $R_\mathrm{n}$. We also assume the temperature is constant over the width of the wire. The electron system looses heat by diffusion along the wire and by electron-phonon interaction. The phonon system looses energy by phonon-diffusion along the wire and escape of phonons from the film to the substrate.

We write a static heat balance equation for both the electron and phonon baths:
\begin{align}
P_\mathrm{DC} + P_\mathrm{diff,e}(x) - P_\mathrm{diff,e}(x+dx) \notag\\\hfill + P_\mathrm{ph-e} - P_\mathrm{e-ph} = 0\\
-P_\mathrm{esc} + P_\mathrm{diff,ph}(x) - P_\mathrm{diff,ph}(x+dx) + \notag\\\hfill P_\mathrm{e-ph} - P_\mathrm{ph-e} = 0
\end{align}

We describe the heat transport by electron and phonon diffusion with Fourier's law:
\begin{eqnarray}
P_\mathrm{diff,e} = -\lambda_\mathrm{e} \frac{dT_\mathrm{e}}{dx}wd,\\
P_\mathrm{diff,ph} = -\lambda_\mathrm{ph}\frac{dT_\mathrm{p}}{dx}wd,
\end{eqnarray}
with $wd$ the wire cross section, and $\lambda_\mathrm{e}$ and $\lambda_\mathrm{ph}$ the electron and phonon heat transfer coefficients:
\begin{eqnarray}
\lambda_\mathrm{e} = \frac{\pi^2}{3}\left(\frac{k_\mathrm{B}}{e}\right)^2\sigma T_\mathrm{e},\\
\lambda_\mathrm{ph} = c_\mathrm{ph}D_\mathrm{ph},
\end{eqnarray}
with $\sigma$ the normal state conductivity, $c_\mathrm{ph}$ the phonon heat capacity, and $D_\mathrm{ph}$ the phonon diffusion coefficient. Assuming two-dimensional phonons, the phonon heat capacity is given by
\begin{equation}
c_\mathrm{ph} \approx 43.3 N_\mathrm{modes} k_\mathrm{B} \left(\frac{T}{\Theta_\mathrm{D}}\right)^2,
\end{equation}
with $N_\mathrm{modes}$ the number of phonon modes per unit volume, and $\Theta_\mathrm{D}$ the Debye temperature. The phonon diffusion coefficient is given by
\begin{equation}
D_\mathrm{ph} = \frac{1}{2}ud,
\end{equation}
with $u$ the speed of sound in the material, and $d$ the film thickness. The electron-phonon coupling is parameterized by a time constant $\tau_\mathrm{e-ph}$. The energy flows are given by
\begin{eqnarray}
P_\mathrm{e-ph} = \frac{c_\mathrm{e}}{\tau_\mathrm{e-ph}}T_\mathrm{e}\cdot w\cdot d\cdot dx,\\
P_\mathrm{ph-e} = \frac{c_\mathrm{ph}}{\tau_\mathrm{ph-e}}T_\mathrm{p}\cdot w\cdot d\cdot dx,
\end{eqnarray}
with $c_\mathrm{e} = \frac{\pi^2}{3}(\frac{k_\mathrm{B}}{e})^2\frac{\sigma}{D_\mathrm{e}}T_\mathrm{e}$ the electronic heat capacity, estimated with the Fermi free electron gas model, with $D_\mathrm{e}$ the electronic diffusion constant. Phonon escape to the substrate, which is assumed to be at the bath temperature, is also parameterized by an escape time $\tau_\mathrm{esc}$:
\begin{equation}
P_\mathrm{esc} = \frac{c_\mathrm{ph}}{\tau_\mathrm{esc}}\left(T_\mathrm{p}-T_\mathrm{b}\right)w\cdot d\cdot dx.
\end{equation}

In this straightforward model, the crucial part is the electron temperature and electronic heat flow. Assuming that electronic inhomogeneities play a role, it may lead to random variations in $T_\mathrm{c}$. However, we find that the resistive transition $R(T)$ curves of the wires are very sharp (inset Fig.~\ref{VI_currcrow}). An alternative source of fluctuations might be contained in the electronic heat conductivity, which would then also be present in the resistivity. In principle, the superconducting state is insensitive to elastic scattering. Variations in elastic scattering can easily be tolerated, while maintaining a uniform $T_\mathrm{c}$\cite{Anderson}. However, in these strongly disordered films the superconducting properties do depend on disorder and a variation of $T_\mathrm{c}$ with position is possible\cite{Driessen}.

For simplicity, we have chosen to model the intrinsic electronic inhomogeneities by letting the critical temperature vary randomly along the length of the wire. At points at which the electron temperature exceeds the local critical temperature, the wire is considered to be in the normal state and Ohmic dissipation is present. At points at which the electron temperature is lower than the critical temperature, the wire is considered to be in the superconducting state and no dissipation is present:

\begin{equation}
P_\mathrm{DC} = \begin{cases}I^2 R_\mathrm{n} \frac{dx}{L}, & \mbox{if } T_\mathrm{e}(x)>T_\mathrm{c}(x)\\ 0, & \mbox{if } T_\mathrm{e}(x)\leq T_\mathrm{c}(x)\end{cases}
\end{equation}

Several parameters play a role but many are known by design (the geometric parameters $w$, $d$, and $L$), or can be measured independently ($D_\mathrm{e}$, $R_\mathrm{n}$, $\sigma$). The following parameters were estimated from literature values for thin NbN films and applied to our NbTiN films:
\begin{itemize}
\item $\tau_\mathrm{e-ph}$. We use an empirical relation $\tau_\mathrm{e-ph} \approx 500T_\mathrm{e}^{-1.6}$ ps, that was reported for NbN thin films\cite{Gousev}. We use the value \emph{at} $T_\mathrm{c}$ and ignore the slight temperature dependence.
\item $\tau_\mathrm{esc}$. The value reported for 3.5~nm NbN on sapphire is $\tau_\mathrm{esc} = 38$~ps\cite{Ilin}.
\item $N_\mathrm{modes}$. This value is taken equal to the atomic density, for NbN this is $N_\mathrm{modes} = 4.8\cdot 10^{28}\ \mathrm{m^{-3}}$\cite{Semenov}.
\item $\Theta_\mathrm{D}$. The two-dimensional Debye temperature is $\Theta_\mathrm{D}^{2D} \approx 0.91\Theta_\mathrm{D}^{3D}$.\cite{Barends} The value of the latter is $\Theta_\mathrm{D}^{3D} = 250\ \mathrm{K}$\cite{Semenov}.
\item $u = 2.3~\mathrm{km/s}$ for NbN\cite{Barends}.
\end{itemize}

\section{Simulated characteristics}

\begin{figure}
\includegraphics{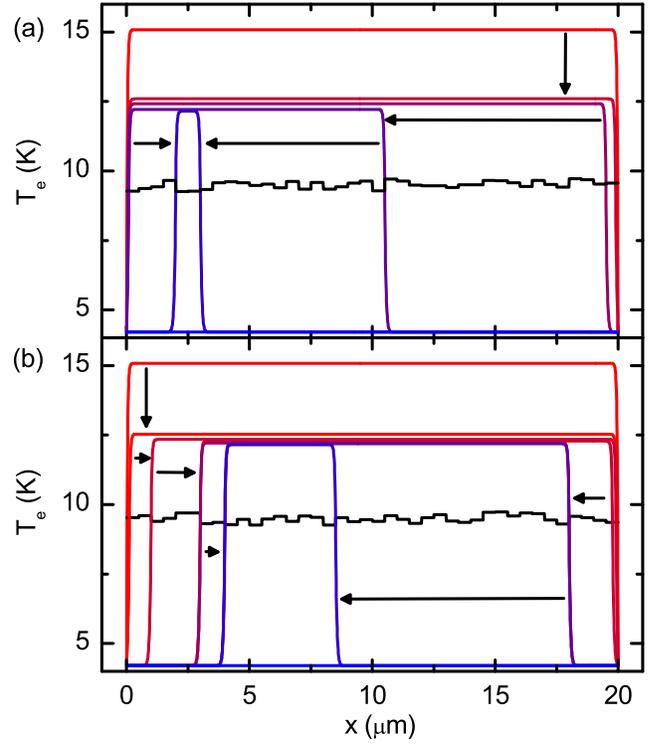}
\caption{Electron temperature profiles along a wire for decreasing current-bias from 122 (red) to 61 $\mu$A (blue). The black line represents the randomly varying critical temperature along the wire. The stepwise transition from normal state to superconducting state can be seen from the temperature profiles in which only a part of the wire has an electron-temperature above the critical temperature.}
\label{TempProf}
\end{figure}

In Fig.~\ref{TempProf}, we show two sets of electron temperature profiles for a 20~$\mu$m long, 1~$\mu$m wide, 8~nm thick, NbTiN wire, simulated using the model described in section \ref{sec:model}. The black line represents the critical temperature, which varies randomly around the experimentally determined value $T_\mathrm{c,avg} = 9.5$~K. The used parameters are $\Delta T_\mathrm{c} = 0.5$~K, where $T_\mathrm{c}(x) = T_\mathrm{c,avg} + \chi\cdot \Delta T_\mathrm{c}$, with $\chi$ a random number between $-\frac{1}{2}$ and $\frac{1}{2}$. The length of the $T_\mathrm{c}$ regions of the random pattern is $500$~nm, a value arbitrarily chosen but suggested by the experiments by Sacepe \emph{et al.}\cite{Sacepe}. In the two panels of Fig.~\ref{TempProf} a different random pattern with the same parameters is chosen, clearly leading to a different approach to the superconducting state. The bias current is decreased from $122$~$\mu$A, at which the entire wire is normal. As the current is lowered, the electron temperature in the wire decreases smoothly, until a stepwise jump occurs in which the superconducting regions at the boundary increase to a new stable position. This continues until finally the entire wire is in the superconducting state.

The phonon-temperature follows the same profile as the electron-temperature. In the normal region it lies between 5.9 and 6.3~K for the currents shown. Because of the low electronic diffusion constant for our films, $D \approx 1~\mathrm{cm^2/s}$, and strong electron-phonon coupling, the energy flow is dominated by energy transfer to the substrate via the phonon system. In other words, the size of the normal regions is always much larger than the thermal healing length in the wire. Therefore, under current-bias, the electron- and phonon-temperatures are constant over the normal region and their value is independent of the size of the normal region.

\begin{figure}
\includegraphics{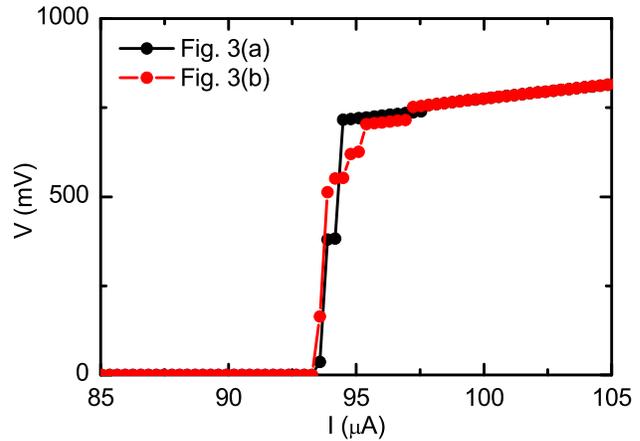}
\caption{Current-voltage characteristics deduced from the simulated electron temperature profiles in Fig.~\ref{TempProf}. Steps in the transition from the normal to the superconducting state are seen for decreasing current-bias. }
\label{VI_sim}
\end{figure}

Fig.~\ref{VI_sim} shows the current-voltage characteristics which results from the electron-temperature profiles in Fig.~\ref{TempProf}. Elements of the wire where the electron temperature is above the critical temperature are taken to be in the normal state, with a resistance, whereas elements of the wire where the electron temperature is below the critical temperature are taken to be in the superconducting state. Jumps in the transition from the normal state to the superconducting state can be seen around 95~$\mu$A. The precise value of the retrapping current differs from the experimentally observed value of $\sim70~\mu$A. However, we are not aiming for a quantitatively accurate prediction of the retrapping current, since a number of thermal parameters are not well-known. We are aiming for possible contributions to the observed steplike features. The exact shape depends on the particular random inhomogeneity pattern of the sample.

Steps in the current-voltage characteristics are also reported in \cite{Elmurodov}, where they are ascribed to phase slip centers. These current-voltage characteristics are however qualitatively different from the ones we observe. In \cite{Elmurodov}, the stepwise transition occurs mainly in the increasing current branch, in which we observe an immediate transition from the superconducting- to the normal-state.

\section{Conclusions}
We observe a persistent stepwise transition from the normal to the superconducting state under decreasing current-bias in NbTiN wires, for all wire geometries. It is argued that the stepwise pattern may be due to disorder fluctuations or electronic inhomogeneities intrinsic to disordered superconductors, modeled by local variations in the critical temperature.

\section*{Acknowledgment}
The authors would like to thank Karl Berggren for stimulating discussions. We acknowledge support from the Foundation for Fundamental Research on Matter (FOM), Microkelvin (No. 228464, Capacities Specific Programme), and the Teradec program of NWO-RFBR.

%\bibliographystyle{IEEEtran}
%\bibliography{IEEEabrv,ASC}

% Generated by IEEEtran.bst, version: 1.13 (2008/09/30)

\end{document}